\documentclass[prd,preprint,showpacs]{revtex4}
\usepackage{amsmath}

\begin{document}
\title{Commutativity of Substitution and Variation in the Actions of Quantum Field Theory}
\author{Zhong Chao Wu}
\email{zcwu2007@yahoo.com} \affiliation{Dept. of Physics, Zhejiang
University of Technology, Hangzhou, Zhejiang 310032, China}

\begin{abstract}
There exists a paradox in quantum field theory: substituting a field
configuration which solves a subset of the field equations into the
action and varying it is not necessarily equivalent to substituting
that configuration into the remaining field equations. We take the
$S^4$ and Freund-Rubin-like instantons as two examples to clarify
the paradox. One must match the specialized configuration field
variables with the corresponding boundary conditions by adding
appropriate Legendre terms to the action. Some comments are made
regarding exceptional degenerate cases.
\end{abstract}
\maketitle

%
%
%
%
%
%
%
%



%
%

%

\section{Introduction}
It is well known in quantum field theory that all field equations
can be derived from an action principle. A field configuration
satisfying the field equations is called a classical solution.

By substituting into the action a field configuration which solves a
subset of the field equations, the action becomes a functional of
the remaining fields. This ``reduced action" would describe the
partial theory in the classical background of the substituted field
configuration. One also expects that varying the reduced action
would result in the same field equations for the remaining variables
as those derived from the original complete action.

However, the situation is not so simple. A paradox exists in the
literature over the fact that substituting a field configuration
which solves a subset of the field equations into the action and
varying it is not necessarily equivalent to substituting the same
field configuration into the remaining field equations [1].

A similar problem had already been encountered earlier in general
relativistic cosmology, where imposition of symmetry on the action
was found not necessarily to give the correct field equations [2],
i.e., imposing symmetry does not necessarily commute with deriving
the field equations from an action principle. The reason for the
failure is that the symmetry imposition may interfere with the
requirement of vanishing variations at the boundary and hence
boundary terms cannot be assumed to vanish.

This problem is very disturbing for quantum field theory. Scientists
have never been lucky enough to have an ultimate theory with the
first try. Instead, they usually consider the known regime as a
classical background, which corresponds to the configuration in the
substitution.  If the above substitution leads to an incorrect
reduced action for the remaining variables, then it is simply
hopeless to make any progress based on the reduced action of the
partial theory. A typical example of this is the consideration of
quantum fields in curved spacetime, in which the metric is assumed
to be a solution of the Einstein equations. In the mid-1970s, even
though little was known about quantum gravity, some very important
progress was made in quantum fields in curved spacetime, the
discovery of the Hawking radiation in the black hole background
being a prime example.

In this paper we use two examples to show that this paradox can be
clarified. These two examples are the $S^4$ instanton and the
Freund-Rubin cosmological models.  The paper can be thought of as a
generalization of a previous paper [3]. We shall discuss the problem
of zero value of the cosmological constant in Section II. In Section
III the 4-sphere model will be studied. Section IV will be devoted
to the Freund-Rubin models. We summarize and conclude this paper in
Section V.

\section{The cosmological constant problem}

The most notable example of this confusion had been associated with
the cosmological constant problem in quantum cosmology [4]. The
issue has been dealt with in the separate publication due to its
special importance [3]. However, it is instructive to give a brief
review here.  To show that the cosmological constant is probably
zero, Hawking considered a cosmological model created from an $S^4$
seed instanton with the Euclidean action [4]
\begin{equation}
I = -\int_M dx^4 g^{1/2}\left ( \frac{1}{16\pi}(R - 2\Lambda_0) -
\frac{1}{8} F^{\mu\nu\rho\sigma}F_{\mu\nu\rho\sigma} \right ),
\end{equation}
where $R$ is the scalar curvature of the space $M$ of 4-metrics,
$\Lambda_0$ represents the contributions of ``the bare cosmological
constant" and the ground states of all matter fields. Here the
contribution from a rank 3 antisymmetric tensor gauge field
$A_{\nu\rho\sigma}$ is singled out. It arises naturally in $N=8$
supergravity in four dimensions [5]. $F$ is the field strength of
$A_{\nu\rho\sigma}$. We use Planck units in which $c = G = k= \hbar
= 1$.

The field configuration [1][4]
\begin{equation}
\sqrt{g}F^{\mu\nu\rho\sigma}=(4!)^{-1/2} \kappa
\epsilon^{\mu\nu\rho\sigma}
\end{equation}
or
\begin{equation}
F_{\mu\nu\rho\sigma}=(4!)^{-1/2}\sqrt{g} \kappa
\epsilon_{\mu\nu\rho\sigma}
\end{equation}
with an arbitrary constant $\kappa$  solves the gauge field equation
\begin{equation}
F^{\mu\nu\rho\sigma}_{\;\;\;\;\;\;\;\; ;\sigma} = 0.
\end{equation}
Substituting this solution (2)--(3) into the action (1), one can see
that the $F^2$ term behaves like an effective cosmological constant
\begin{equation}
\Lambda_{eff}= \pi \kappa^2,
\end{equation}
and the total cosmological constant is $\Lambda_{total} = \Lambda_0
+ \Lambda_{eff}$. For $\Lambda_{total}$, the radius of $S_4$ is
$(3/\Lambda_{total})^{1/2}$, and the action is $-
3\pi/\Lambda_{total}$. Here it is assumed that $\Lambda_{total}$ is
positive. The action is the negative of entropy of the created de
Sitter spacetime. The relative creation probability of the universe
is the exponential of the negative of the action [6]. It follows
that the most probable configurations will be those with very small
values of $\Lambda_{total}$, and nature will automatically select
the right value of $\kappa$ for this. Therefore, Hawking concluded
[4]: ``the cosmological constant is probably zero."

However, Duff showed that after substituting the configuration into
the Einstein equation, which is derived from the complete action
(1), the Einstein equation reads
\begin{equation}
G^{\mu\nu} = \pi \kappa^2g^{\mu\nu} - \Lambda_0g^{\mu\nu}.
\end{equation}

Comparing the reduced action and substituted field equation (6), one
finds the total cosmological constant appearing in the action is not
the same as that appearing in the field equation [1]. What we truly
observe is in (6)!

This dilemma can be resolved by choosing the right representation
for the wave function of the universe at the equator of the
instanton, where the quantum transition from the Euclidean regime
$S^4$ to the Lorentzian regime $dS^4$ occurs. The instanton should
not be simply considered as  $S^4$, it must be considered as a union
of a southern hemisphere joined to its time reversal, the northern
hemisphere. The action (1) corresponds to the boundary condition
that $A_{\nu\rho\sigma}$ is given at the equator between the two
hemispheres for the creation probability calculation. However,
$A_{\nu\rho\sigma}$ is the wrong representation, which suffers a
discontinuity across the equator. Here, we have set
$A_{\nu\rho\sigma}$ to be regular in each hemisphere. Therefore one
has to use the right representation, its conjugate variable
$\sqrt{g}F^{\mu\nu\rho\sigma}$, which is continuous across the
equator. The representation transformation is carried out by adding
a Legendre term to the action (1)
\begin{equation}
I_{Legendre}= -\int_{\Sigma_{S+N}} dS_\mu
A_{\nu\rho\sigma}F^{\mu\nu\rho\sigma},
\end{equation}
where $\Sigma_{S+N}$ denotes the two equator boundaries for both the
southern and northern hemispheres.

The total action is
\[
I_{total} = -\int_M dx^4 g^{1/2}\left ( \frac{1}{16\pi}(R -
2\Lambda_0) - \frac{1}{8} F^{\mu\nu\rho\sigma}F_{\mu\nu\rho\sigma}
\right )-\int_{\Sigma_{S+N}} dS_\mu
A_{\nu\rho\sigma}F^{\mu\nu\rho\sigma}
\]
\begin{equation}
= -\int_M dx^4 g^{1/2}\left ( \frac{1}{16\pi}(R - 2\Lambda_0) +
\frac{1}{8} F^{\mu\nu\rho\sigma}F_{\mu\nu\rho\sigma} \right ),
\end{equation}
where the second equality is obtained by taking divergence of the
Legendre term and using the gauge field equation (4).

Substituting (2)--(3) into (8) yields
\begin{equation}
I_{total} = -\int_M dx^4\left (\frac{g^{1/2}}{16\pi}(R - 2\Lambda_0)
+ \frac{1}{8}g^{1/2} \kappa^2 \right ).
\end{equation}

Apparently, varying the action (9) with respect to the gravitational
field will result in the same Einstein equation (6). Therefore,
Duff's dilemma about the cosmological constant is dispelled and
Hawking's argument is completely proven [3].

Two decades after the publication of Hawking's paper, many people
believe that the cosmological constant is not zero. Therefore,
Duff's paradox has remained long forgotten.

\section{The 4-sphere model}

However, the paradox as a whole has not been resolved. If one is
dealing with the simple $S^4$ instanton without any boundary,
instead of the southern hemisphere joined to the northern hemisphere
of $S^4$ in the above quantum creation scenario, then it seems that
there is no reason to reject the representation $A_{\nu\rho\sigma}$.
Even for the representation $\sqrt{g}F^{\mu\nu\rho\sigma}$ it seems
that no Legendre term has to be added since there seems to be no
boundary in the $S^4$ instanton model. Therefore, the dilemma still
persists.

Now, let us study this simple $S^4$ model with the action (1). As
one derives the Einstein and gauge field equations from the action,
it is implicitly assumed that the metric $g_{\mu\nu}$ and the gauge
potential $A_{\mu\nu\rho}$ are independent variables. In varying the
action, one imposes the condition that $A_{\mu\nu\rho}$ is fixed at
the boundary.

There is no boundary for the metric of the $S^4$ model. People may
wonder, why we should bother with the boundary value problem here.
The point is that the gauge potential is involved. Since one has to
use at least two patches to cover the manifold for the gauge
potential, the Legendre term should arisen from the boundary between
the patches under some circumstances. In fact, here we are not
particularly interested in varying the action with respect to the
gauge field, it has been done and the solution (2) was already found
for (4). The key point is that the boundary term for the gauge field
is also crucial for our main motivation, the next step: varying the
reduced action with respect to the remaining variable, the
gravitational field.

For convenience, the reduced action obtained from substituting
(2)--(3) into (1) can be written as follows
\begin{equation}
I = -\int_M dx^4\left ( \frac{g^{1/2}}{16\pi}(R - 2\Lambda_0)
-\frac{1}{8} {g}^{-1/2}\hat{ \kappa }^2 \right ),
\end{equation}
where $\hat{\kappa} \equiv g^{1/2} \kappa$. Now one can vary the
reduced action (10) with respect to the gravitational field under
the condition that $A_{\mu\nu\rho}$ is fixed. The condition is
equivalent to  $F_{\mu\nu\rho\sigma}$ being fixed, i.e,
$\hat{\kappa}$ is fixed using (3). Here, for simplicity, the
minisuperspace ansatz is imposed. This results in the Einstein
equation (6), as expected.

One can equally vary the reduced action with respect to the
gravitational field under the condition that the conjugate variable
$\sqrt{g}F^{\mu\nu\rho\sigma} $ is fixed. It is noted that the
variables $ \sqrt{g}F^{\mu\nu\rho\sigma}$ and $F_{\mu\nu\rho\sigma}$
are not equivalent in the presence of a gravitational field. Here,
we are using $g_{\mu\nu}$ and $\sqrt{g}F^{\mu\nu\rho\sigma}$ as
independent variables. For consistency, one has to change the
boundary condition correspondingly, that is, add a proper Legendre
term to the action.

As emphasized earlier, for the gauge potential $A_{\mu\nu\rho}$, it
is impossible to cover the manifold $S^4$ by one patch. We can
decompose the manifold into two parts: a small ball $B$ and $S^4 -
B$, which we denote as $A$. Following the above procedures, the
Legendre term is
\begin{equation}
I_{Legendre}= -\int_{\partial A + \partial B} dS_\mu
A_{\nu\rho\sigma}F^{\mu\nu\rho\sigma}.
\end{equation}
One can use the gauge field equation (4) to reduce the Legendre
surface term to a divergence term as in (8). Here, we set
$A_{\nu\rho\sigma}$ to be regular in $A$ and $B$, respectively.
There is an unavoidable discontinuity of the potential across the
boundary between them. It turns out that the contributions of the
Legendre terms are $-\kappa^2 V_B/4$ and $-\kappa^2 V_A/4$, where
$V_B$ and $V_A$ are their volumes.  The total contribution of these
two Legendre terms for the instanton is $-\kappa^2V_{S^4}/4$.

Now, we can let $B$ shrink to a point, then the contributions of $B$
to both the volume and boundary terms of the action tend to zero,
and we can discard them. In any case, the total action for $S^4$ is
(8)--(9). This is the same as that in the model for the cosmological
constant problem. The Einstein equation can be derived from the
reduced action in the same way.

Indeed, a byproduct of the above argument is that we have proven
that it is impossible to regularly express $A_{\mu\nu\rho}$ for the
whole $S^4$ in one piece or a single gauge. Otherwise, the Legendre
term would vanish after shrinking, and this would lead to a
contradiction. Therefore, even though $S^4$ has no boundary, the
inevitable singularity of $A_{\mu\nu\rho}$ leads to the Legendre
term.

In the following, for simplicity, we use gauge freedom to force the
singularity to be located at the center of $B$. This is also true
for the $S^n (n \ge 2)$ model with a rank $(n-1)$ antisymmetric
tensor Abelian gauge field.

In summary, what we did is to cover $S^4$ by the two patches: $A$
and $B$. For the gauge potential, one can only cover it by at least
two arbitrary patches. If we use the representation
$A_{\mu\nu\rho}$, we have to tolerate the discontinuities across the
boundaries of these patches. If we prefer using the representation
$\sqrt{g}F^{\mu\nu\rho\sigma} $ to avoid these discontinuities, then
we have to add the Legendre terms, whose total contribution takes
the same form as (11),  where $\partial A + \partial B$ is replaced
by all these boundaries. The total action remains the same as (8).
Our choice is just for simplicity. There is no topology change when
one lets $B$  shrink to a point. The same arguments apply to the
next model. However, for the wave function of the universe only the
representation $\sqrt{g}F^{\mu\nu\rho\sigma} $ can be used.


\section{The Freund-Rubin models}

Next turn to a more complicated model, the Freund-Rubin-like
instanton [7]. This model has been used to investigate
dimensionality of spacetime in quantum cosmology [8]. The spacetime
manifold $M$ is described by a product of two spheres $S^s \times
S^{n-s}$. Apparently, the $S^4$ model is a special case with $n = s
= 4$. The seed for quantum creation of a black hole or a
codimension-2 braneworld [9] is an instanton of topology $S^s \times
S^2$, where $S^2$ can be a distorted sphere with some conical
singularities. It is trivial to generalize our discussion to this
case. For simplicity, we shall only consider the Freund-Rubin-like
instanton, and the action takes the form
\begin{equation}
I = -\int_M dx^{n}g^{1/2}\left (\frac{1}{16\pi}(R -2\Lambda )-
\frac{1}{2s} F^2 \right ) - \frac{1}{8\pi}\int_{\partial
M}dx^{n-1}h^{1/2} K,
\end{equation}
where $R$ is the scalar curvature of $M$, $K$ is the extrinsic
curvature of the boundary $\partial M$ with induced metric $h_{ij}$,
$\Lambda$ is the cosmological constant, and $F$ is the field
strength of a rank $-s-1$ antisymmetric Abelian tensor
$A_{\alpha_1\cdots\alpha_{s-1}}$. We shall use the ansatz that all
indices of nonvanishing $F$ components should reside in $S^s$. The
Euclidean action is obtained via an analytic continuation from the
Lorentzian action. There is some overall sign ambiguity in the
action. The ambiguity can be eliminated by the following
consideration. In order for the primordial fluctuations to take the
ground states allowed by the Heisenberg uncertainty principle [10],
the term associated with the scalar curvature of the external factor
spacetime must be negative. But this ambiguity will not affect our
discussion.

Similarly, the gauge field equation is
\begin{equation}
F^{\alpha_1\cdots\alpha_s}_{\;\;\;\;\;\;\;\; ;\alpha_s} = 0 \;\;\;
(1\leq \alpha_i \leq n)
\end{equation}
and under the ansatz the solutions must take the form
\begin{equation}
F^{\alpha_1\cdots\alpha_s } = \left\{  \begin{array}{cc}
(s!g)^{-1/2} \kappa \epsilon^{\alpha_1\cdots\alpha_s}&(1\leq
\alpha_i \leq s) \\
0& \mbox{otherwise},
\end{array}
\right.
\end{equation}
where $g$ is the metric determinant of the product space, and
$\kappa$  is an arbitrary constant. It follows that
\begin{equation}
F_{\alpha_1\cdots\alpha_{s}} =\left\{  \begin{array}{cc}
{(s!g_{n-s})}^{-1/2}{(g_s)}^{1/2}\kappa\epsilon_{\alpha_1\cdots\alpha_{s}}\;\;\;(1\leq
\alpha_i \leq s)\\
0&\mbox{otherwise},
\end{array}
\right.
\end{equation}
where $g_s$ and $g_{n-s}$ are the metric determinants of the factor
spaces.

The Einstein equation is
\begin{equation}
G^{\mu\nu} = 8\pi \theta^{\mu\nu} -\Lambda g^{\mu\nu},
\end{equation}
where the stress tensor is
\begin{equation}
\theta^{\mu\nu} =
F_{\alpha_1\cdots\alpha_{s-1}}^{\;\;\;\;\;\;\;\mu}F^{\alpha_1\cdots\alpha_{s-1}\nu}
- \frac{1}{2s}
F_{\alpha_1\cdots\alpha_s}F^{\alpha_1\cdots\alpha_s}g^{\mu\nu}.
\end{equation}

From the Einstein equation one can derive the scalar curvatures of
the factor spaces
\begin{equation}
R_s = \frac{(n-s-1)8\pi \bar{\kappa}^2}{n-2} +
\frac{2s\Lambda}{n-2},
\end{equation}
\begin{equation}
R_{n-s} = -\frac{(s-1)(n-s)8\pi \bar{\kappa}^2}{s(n-2)}+
\frac{2(n-s)\Lambda}{n-2},
\end{equation}
where $\bar{\kappa} \equiv (g_{n-s})^{-1/2}\kappa$, and it is noted
that $\bar{\kappa}$ depends on the metric of the factor space
$S^{n-s}$. From these curvatures one can derive the radii of the
factor spheres.

Now we substitute the gauge field configuration (14)--(15) into the
action (12), then the reduced action is
\begin{equation}
I= -\frac{1}{16\pi} \int_M
dx^n\left(({g_s})^{1/2}{(g_{n-s})}^{1/2}(R - 2\Lambda) -
\frac{8\pi}{s}({g_s})^{-1/2}({g_{n-s}})^{1/2} \hat{\kappa}^2 \right
).
\end{equation}
where $\hat{\kappa} \equiv (g_{n-s})^{-{1/2}}(g_s)^{1/2}\kappa$.

One can easily derive the Einstein equation  from (20) without a
boundary term, considering $A_{\alpha_1\cdots\alpha_{s-1}}$ or
$\hat{\kappa}$ as the given independent variable, as we did in the
$S^4$ model,
\begin{equation}
G^{\mu\nu} = -\Lambda g^{\mu\nu}
+\frac{4\pi\bar{\kappa}^2g^{\mu\nu}}{s}, (1 \leq \mu, \nu \leq s),
\end{equation}
\begin{equation}
G^{\mu\nu} =-\Lambda g^{\mu\nu} -
\frac{4\pi\bar{\kappa}^2g^{\mu\nu}}{s}, (s+1 \leq \mu, \nu \leq n),
\end{equation}
which is exactly the same as that derived from the complete action
(12). Eqs. (18)--(19) are implied by (21)--(22), of course.

One can also derive these from (20) with the Legendre term,
considering  $\sqrt{g} F^{\alpha_1\cdots\alpha_s}$ as the
independent variable. As usual, the corresponding Legendre term is
\begin{equation}
I_{Legendre} = -\int_{(\partial \bar{S}^s)\times
S^{n-s}}dS_{\alpha_1}
A_{\alpha_2\cdots\alpha_{s}}F^{\alpha_1\cdots\alpha_{s}},
\end{equation}
where $\bar{S}^s$ denotes $S^s$ minus a point where the singularity
of $A_{\alpha_1\cdots\alpha_{s-1}}$ is located, as in the $S^4$
model. The integral is over (the $A$ singularity in $S^s $)$\times
S^{n-s}$ (in the sense of shrinking $B$ to its center).

Using the gauge field equation (13), the Legendre term can be
written
\begin{equation}
I_{Legendre} = -\int_{S^s \times S^{n-s}}dx^n g^{1/2}\frac{ F^2}{s},
\end{equation}
where
\begin{equation}
F^2 =  F_{\alpha_1\cdots\alpha_{s}}
F^{\alpha_1\cdots\alpha_{s}}=\left ({g_{n-s}}\right )^{-1}\kappa^2.
\end{equation}
The total action is
\begin{equation}
I_{total}=- \frac{1}{16\pi} \int_M dx^n g^{1/2}\left(R - 2\Lambda +
\frac{8\pi}{s} F^2 \right ).
\end{equation}
Using (25), one can recast (26) into the form
\begin{equation}
I_{total}= -\frac{1}{16\pi} \int_M
dx^n\left(({g_s})^{1/2}{(g_{n-s})}^{1/2}(R - 2\Lambda) +
\frac{8\pi}{s}({g_s})^{1/2}({g_{n-s}})^{-1/2} \kappa^2 \right ).
\end{equation}

Varying the reduced total action (27) with respect to the metric for
the given independent $\kappa$, one can also derive the same
Einstein equation (21)--(22). Nature is always self-consistent!

\section{Conclusion}

For a general case, if an original solution is extremal to the
original action, then the solution obtained from the reduced action
should be extremal too, as long as we supplement the action with an
appropriate boundary term. However, if the complete solution is
non-extremal stationary to the original action, one has to be
cautious, the solution that solves the subset of the equations might
trace an orbit with constant action value, then the whole orbit may
be the solution set for the reduced action and the original solution
for the complete action would be one in the solution set after the
substitution. We call this case degenerate. Since the Euclidean
action of gravitational field is not positive definite, so one has
to pay close attention to it. However, this issue does not affect
our models due to our choice of variables.

In summary, we have used two models to show that substituting the
configuration into the action and then varying it is equivalent to
substituting it into the field equation, except for the above
degenerate case. The key point is to identify the variables
substituted and provide the corresponding boundary conditions, which
are implemented by adding to the action the proper Legendre terms
needed. In general, the choice of independent variables is made
based on convenience. The origin of the inconsistency in the earlier
literature is the mismatch of the substituted fields and their
boundary conditions. It is believed that as long as we keep this in
mind, everything is consistent and the reduced action is viable for
the partial theory.

\section*{Acknowledgement}
I would like to thank G.W. Gibbons and D. Lohiya for helpful
discussions. This work is supported by NSFC No.10703005 and
No.10775119.

%

\end{document}